\documentclass{raa}          

\usepackage{graphicx,times,epsf}             %for PS/EPS graphics inclusion, new
\usepackage{amsmath}
\usepackage{makecell}

\begin{document}
   \title{On the central symmetry of the circumstellar envelope of RS\,Cnc \thanks{Based on observations carried out with the IRAM Plateau de Bure Interferometer and the IRAM 30 m telescope. IRAM is supported by INSU/CNRS (France), MPG (Germany) and IGN (Spain).}}
%   \subtitle{I. Place Your Subtitle Here}
   \volnopage{Vol.0 (200x) No.0, 000--000}      %%preserved for Editor. DOn't remove!
   \setcounter{page}{1}          %%starting page, preserved for Editor. DOn't remove!  
   \author{Pham Tuyet Nhung\inst{1}
          \and
          Do Thi Hoai\inst{1,2}
          \and
          Jan Martin Winters\inst{3}
          \and
          Pierre Darriulat\inst{1}
          \and  \\        
          Eric G\'erard\inst{4}
          \and
          Thibaut Le Bertre\inst{2} 
     }     
   \institute{VATLY, INST, 179, Hoang Quoc Viet, Cau Giay, Hanoi, Vietnam \and
             LERMA, UMR 8112, CNRS \& Observatoire de Paris, 61 av. de l'Observatoire, F-75014 Paris, France \and
             IRAM, 300 rue de la Piscine, Domaine Universitaire, F-38406 St. Martin d'H\`eres, France \and
             GEPI, UMR 8111, CNRS \& Observatoire de Paris, 5 Place J. Janssen, F-92195 Meudon Cedex, France 
             }
   \date{}
   \abstract{We present a phenomenological study of CO(1-0) and CO(2-1) emission from the circumstellar envelope (CSE) of the Asymptotic Giant Branch (AGB) star RS\,Cnc. It reveals departures from central symmetry that turn out to be efficient tools for the exploration of some of the CSE properties. We use a wind model including a bipolar flow with a typical wind velocity of $\sim$8 km\,s$^{-1}$ decreasing to $\sim$2 km\,s$^{-1}$ near the equator to describe Doppler velocity spectral maps obtained by merging data collected at the IRAM Plateau de Bure Interferometer and Pico Veleta single dish radio telescope. Parameters describing the wind morphology and kinematics are obtained, together with the radial dependence of the gas temperature in the domain of the circumstellar envelope probed by the CO observations. Significant north-south central asymmetries are revealed by the analysis, which we quantify using a simple phenomenological description. The origin of such asymmetries is unclear.
   }
   \authorrunning{P. T. Nhung et al.}            %author_head in even pages
   \titlerunning{}  % title_head in odd pages
   \maketitle
%% The author head (on even pages) and the title head (on odd pages) will be
%% automatically extracted from \author{} and \title{}. Whenever the title is too long,
%% you will be asked to supply a shorter one by inserting either \authorrunning{} or
%% \titlerunning{} before \maketitle. Anyway, you can specify your own heads.
%%
%%
%% Note: In the following text body of your manuscript, please note several differences from
%%       other major journals:
%% (1) \subsection{Please Capitalize the First Letter of Each Notional Word in Subsection Title}
%% (2) Please Capitalize the First Letter of Each Notional Word in all tables' captions

%
%________________________________________________ sections below

%(Clemens~\cite{clem85})  
\section{Introduction}

RS Cnc, a semi-regular variable star presently in the thermally pulsing phase of the Asymptotic Giant Branch (AGB), is one of the best targets to observe the mass loss process undergone by this kind of objects, thanks to its proximity (d $\sim$140 pc). Libert et al.~(\cite{libe}) have obtained CO(1-0) and CO(2-1) interferometric maps of its circumstellar environment, from which they infer an axi-symmetric bipolar outflow. Recently Hoai et al.~(\cite{hoai}) revisited the source with better quality data and developed a model, in which the velocity and the flux of matter increase with latitude from an equatorial plane to a polar axis, which reproduces fairly well the observed spectral maps. It is interesting to investigate this phenomenon in those AGB sources where it is present and to understand its origin, as it may hold clues to the wide spread axi-symmetry observed in post-AGB and PN nebulae (e.g. Sahai et al.~\cite{saha}). The presence of a magnetic field, stellar rotation, and/or binarity have been invoked, although no consensus has emerged. The interplay of these effects may also have an important role in the phenomenon of mass loss itself. It is noteworthy that RS Cnc exhibits composite CO line profiles, indicating two regimes of mass-loss and velocity. Besides RS Cnc, there are other AGB stars where a bipolar outflow is suspected, like EP Aqr (Winters et al.~\cite{wint07}) or X Her (Castro-Carrizo et al.~\cite{cast}), and which also show composite CO line profiles (e.g., Knapp et al.~\cite{knap}, Winters et al.~\cite{wint03}) so that RS Cnc may not be an isolated case.

The present work uses the same data as Hoai et al.~(\cite{hoai}) to study the central symmetry of the gas distribution and kinematics, namely to which extent diametrically opposite gas volumes (with respect to the centre of the star) have equal densities and temperatures and opposite velocities. The question is of relevance to the late evolution of AGB stars that are known to commonly evolve into Planetary Nebulae having strongly asymmetric morphologies (for recent reviews, see e.g. Habing and Olofsson~\cite{habi}, Herwig~\cite{herw}, Marengo~\cite{mare}, Bujarrabal~\cite{buja1}, Castro-Carrizo et al.~\cite{cast}, Pascoli and Lahoche~\cite{pasc}, Zhao-Geisler~\cite{zhao}, Amiri~\cite{amir}, Geise~\cite{geis}, Lagadec~\cite{laga} and Bujarrabal et al.~\cite{buja2} and references therein). 

The idea that the wind accelerates at the end of the AGB phase from velocities in the 10 km\,s$^{-1}$ ballpark to velocities in the 100 km\,s$^{-1}$ ballpark, and that the new superwind interacts with the old wind to eject huge masses of gas is well accepted (Kwok et al.~\cite{kwok1} \&~\cite{kwok2} and references therein). However, the mechanism that triggers such superwind is unclear and the reason for a deviation from spherical symmetry during an earlier phase is not well understood. Possible causes include binarity, magnetic fields, large convection zones or clumpiness of the wind. AGB stars, Post-AGB stars and young Planetary Nebulae have been observed at infrared, mid-infrared, far-infrared, millimeter and submillimeter wavelengths with morphologies including, in both their gas and dust contents, a bipolar outflow and/or an equatorial disk or torus perpendicular to it. Searching for a possible central asymmetry in the circumstellar envelope of RS\,Cnc provides useful information in this respect.

\section{Signatures of central symmetry}
If the star velocity field \textit{\textbf{V}} and density $n$ distributions are symmetric with respect to the centre of the star (central symmetry), the following relations must be obeyed: $n(x,y,z)=n(-x,-y,-z)$ and \textit{\textbf{V}}$(x,y,z)=-$\textit{\textbf{V}}$(-x,-y,-z)$. We use coordinates such that $x$ points away from the observer on the line of sight, $y$ points East and $z$ points North. To the extent that emission can be considered as optically thin (the effect of absorption is discussed in Section 5), the flux density measured at velocity $v$ (red shifted when positive) on a line of sight $(y,z)$ is, up to a constant factor,
\begin{equation}
 F(v,y,z)=\int{n(x,y,z)exp(-\tfrac{1}{2}[V_x(x,y,z)-v]^2/\xi^2)dx}
\end{equation}
where $V_x$ is the $x$ component of the velocity and $\xi$ is a smearing parameter defining the velocity resolution.
\begin{equation}
\begin{array}{l}
F(-v,-y,-z)=\int{n(x,-y,-z)exp(-\tfrac{1}{2}[V_x(x,-y,-z)+v]^2/\xi^2)dx} \\
=\int{n(-x,-y,-z)exp(-\tfrac{1}{2}[V_x(-x,-y,-z)+v]^2/\xi^2)dx}
\end{array}
\end{equation}
In the case of central symmetry,
\begin{equation}
F(-v,-y,-z)=\int{n(x,y,z)exp(-\tfrac{1}{2}[-V_x(x,y,z)+v]^2/\xi^2)dx}=F(v,y,z)
\end{equation}

It is convenient to introduce the symmetric and antisymmetric components of each pair of diametrically opposite spectra, respectively labelled $dir$ (for direct spectrum) and $mir$ (for mirror spectrum), defined as follows. 

The sum of a spectrum measured at $(y,z)$ and that measured at $(-y,-z)$ reads $\Sigma_{dir}(v,y,z)=F(v,y,z)+F(v,-y,-z)$. In the case of central symmetry, $\Sigma_{dir}(v,y,z)=F(v,y,z)+F(-v,y,z)=\Sigma_{dir}(-v,y,z)$. The difference is $\Delta_{dir}(v,y,z)=F(v,y,z)-F(v,-y,-z)$. In the case of central symmetry, $\Delta_{dir}(v,y,z)=F(v,y,z)-F(-v,y,z)=-\Delta_{dir}(-v,y,z)$.

The difference between a spectrum at $(y,z)$ and the mirror spectrum at $(-y,-z)$ is $\Delta_{mir}(v,y,z)=F(v,y,z)-F(-v,-y,-z)$. It must cancel in the case of central symmetry. The sum reads $\Sigma_{mir}(v,y,z)=F(v,y,z)+F(-v,-y,-z)$. By construction, $\Sigma_{mir}(v,y,z)=\Sigma_{mir}(-v,-y,-z)$.

The three relations $\Delta_{mir}=0$, $\Sigma_{dir}(v)=\Sigma_{dir}(-v)$ and $\Delta_{dir}(v)=-\Delta_{dir}(-v)$ are therefore signatures of central symmetry for each pair of diametrically opposite spectra. As mentioned in the introduction, central symmetry is meant to include velocity, density and temperature. Indeed, it is not sufficient for the density to be centrally symmetric for the flux associated with different rotational levels of the CO molecule to obey central symmetry.

\section{Observations}
The observations used in the present study are from the Plateau de Bure Interferometer and the Pico Veleta 30 m telescope and have been presented in Hoai et al.~(\cite{hoai}) together with a description of data collection and reduction that does not need to be repeated here. They consist in two sets of 13$\times$13 continuum subtracted velocity spectra, one for CO(1-0) and one for CO(2-1), covering 18.2$\arcsec$$\times$18.2$\arcsec$ (1$\arcsec$ corresponds to $\sim$140 a.u). Each spectrum covers 1.4$\arcsec$$\times$1.4$\arcsec$ and includes 101 velocity bins of 0.2 km\,s$^{-1}$ each. In the present work we generally restrict the analysis to the 49 (7$\times$7) central spectra covering 9.8$\arcsec$$\times$9.8$\arcsec$. The synthesized circular beams are Gaussian with a full width at half maximum of 1.2$\arcsec$.  

The evaluation of the four quantities $\Sigma_{mir}$, $\Delta_{mir}$, $\Sigma_{dir}$ and $\Delta_{dir}$ implies the knowledge of the star velocity and position. In a first phase, the analysis used as origin of the declination and right ascension scales, on which the spectral maps were centred, the values (30$^\circ$57$\arcmin$47.3$\arcsec$ and 9h 10min 38.80s) at epoch 2000.0 measured from Hipparcos data (van Leeuwen,~\cite{vanl}). It used as star Doppler velocity, on which the velocity spectra themselves were centred, that obtained from a fit (Hoai et al.~\cite{hoai}) to the observed data of a model including a bipolar outflow with a typical wind velocity of $\sim$8 km\,s$^{-1}$ decreasing to $\sim$2 km\,s$^{-1}$  near the equator (referred to as standard model in what follows). However, the asymmetries introduced by the proper motion of the star between year 2000 and the times at which observations were made (2004-2005 and 2011) were found to introduce significant central asymmetries and the data were reprocessed using the position of the star at the time of observations as centre of the spectral maps. Moreover, this preliminary study of the central symmetry also revealed an offset of 0.5 km\,s$^{-1}$ between the CO(1-0) and CO(2-1) data that was soon understood as being caused by a small imprecision of the CO(1-0) emission frequency used in data reduction. This was also corrected when reprocessing the data and the observations used in the present work are therefore expected to be bias-free from the point of view of central symmetry. The standard model is centrally symmetric with respect to the star, except for self-absorption in the CO lines, which, however, induces negligible asymmetries. 

\section{Description of CO(1-0) and CO(2-1) emissions using a centrally symmetric model}

\begin{figure}
   \centering
   \includegraphics[width=0.9\textwidth]{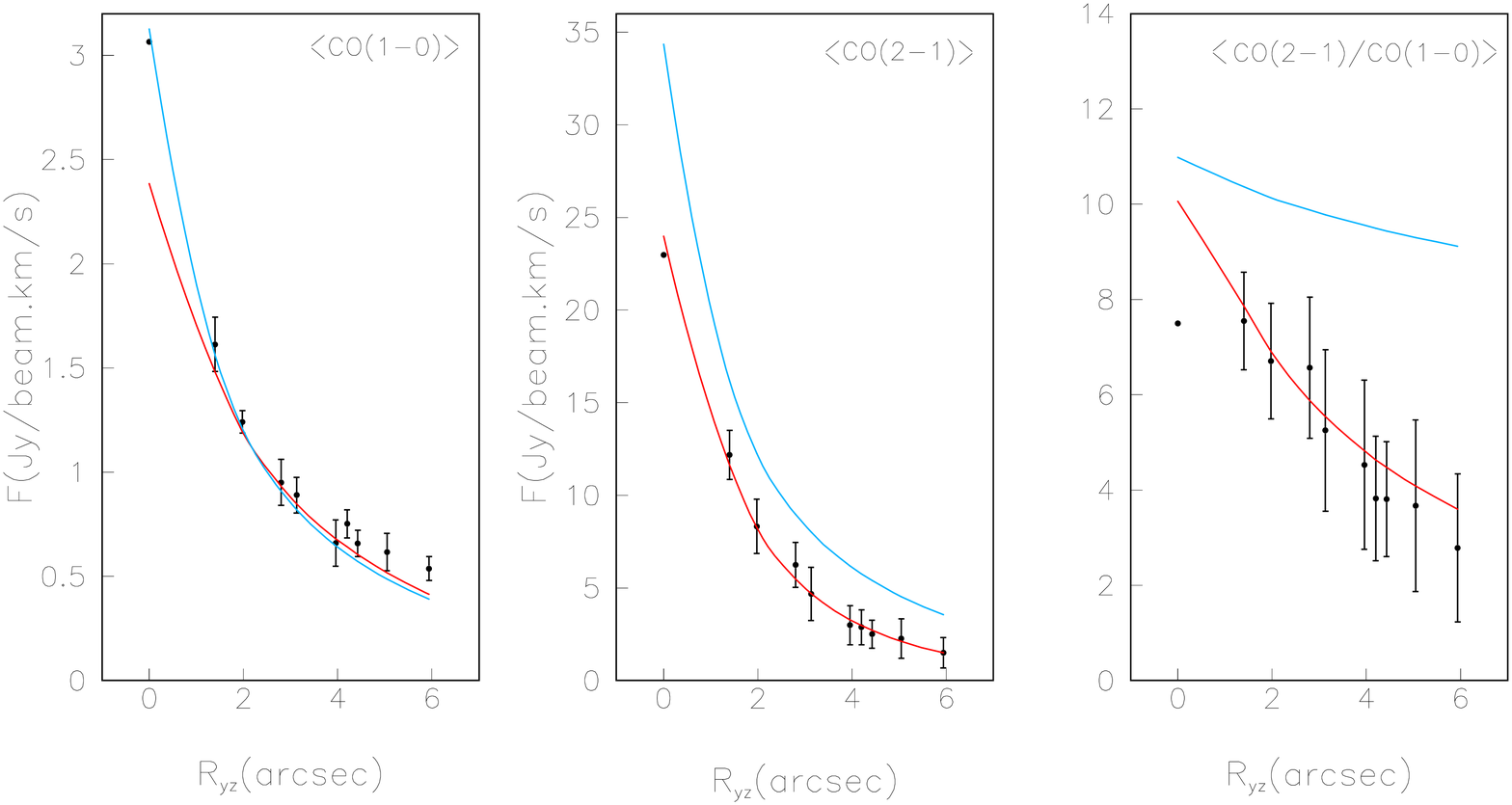}
   \caption{$R_{yz}$ distributions of \mbox{$\langle$\textit{CO(1-0)}$\rangle$} (\textit{left panel}) and \mbox{$\langle$\textit{CO(2-1)}$\rangle$} (\textit{middle panel}) as observed (\textit{black symbols}) and obtained from the standard model in its present version (\textit{red curves}) or in the version used in Hoai et al.~(\cite{hoai}) (\textit{cyan curves}).  \textit{Right panel}: $R_{yz}$ dependence of the averaged ratio \mbox{$\langle$\textit{CO(2-1)/CO(1-0)}$\rangle$} as observed (\textit{black symbols}) and obtained from the model (\textit{red and cyan curves}). The best fit results of the standard model (shown here) are essentially identical to those of its modified asymmetric version. The error bars show root mean square deviations with respect to the mean (not defined at $r=$\,0 where a single cell contributes).}
       \label{Fig1}
 \end{figure}

A major difference between the CO(1-0) and CO(2-1) observations is the larger extension of the former with respect to the latter (Libert et al.~\cite{libe}, Figure 3). Figure~\ref{Fig1} compares the $R_{yz}$ distributions observed for CO(1-0) and CO(2-1) emissions with the best fit results of the standard model (see below). Here, $R_{yz}$ is the angular distance in the plane of the sky between the centre of the star and the central line of sight of each of the 49 velocity spectra (the impact parameter). For each value of $R_{yz}$ ( 0$\arcsec$, 1.4$\arcsec$, 2.0$\arcsec$, 2.8$\arcsec$, etc.) there are a number of spectra (1, 4 or 8) over which the flux densities are averaged to obtain \mbox{$\langle$\textit{CO(1-0)}$\rangle$} and \mbox{$\langle$\textit{CO(2-1)}$\rangle$}. The error bars are the root mean square deviations from the mean for each of the $R_{yz}$ values. They give an upper limit of the uncertainties attached to the data, which we estimate not to exceed 5$\%$.
%Fig 1

In the standard model, the assumption that the winds are stationary implies that the $r$-dependence of the gas density is uniquely determined by the velocity gradients and does not leave freedom to fit separately their contributions to the velocity and radial distributions. This is an obvious oversimplification of reality. Indeed, the mass loss rates and wind velocities are parameterized at each value of the sine of the star latitude, $\gamma$,  in the form $\dot{M}(\gamma)=\dot{M}_1F(\gamma)+\dot{M}_2$ and $V(\gamma,r)=V_1F(\gamma)(1-\lambda_1e^{-r/2.5\arcsec})+V_2(1-\lambda_2e^{-r/2.5\arcsec})$. Here, $\dot{M}_1$, $\dot{M}_2$, $V_1$, $V_2$, $\lambda_1$ and $\lambda_2$ are six adjustable parameters and the function $F(\gamma)$ uses Gaussian profiles centred at the poles, $F(\gamma)=exp(-\tfrac{1}{2}[\gamma-1]^2/\sigma^2)+exp(-\tfrac{1}{2}[\gamma+1]^2/\sigma^2)$, where a seventh parameter $\sigma$ measures the angular aperture of the bipolar flow. Two additional angles, angle of inclination of the polar axis over the plane of sky (AI) and position angle of the projection of this axis over the plane of the sky (PA), are necessary to fix its orientation in space making a total of nine adjustable parameters. The density $n(\gamma,r)$ is then defined at any point from the relation $\dot{M}(\gamma)=4\pi r^2V(\gamma,r)n(\gamma,r)$ and its $r$-dependence is exclusively controlled by the parameters $\lambda_1$ and $\lambda_2$ that define the velocity gradients. 

The wind is supposed to be purely radial, free of turbulences and in local thermal equilibrium. Moreover, it is supposed to have been in such a regime for long enough a time, such that the radial extension of the gas volume, in the 10$\arcsec$ ball park, is governed exclusively by the UV dissociation of the CO molecules from the ISM and does not keep any track of the star history. At a velocity of 2 km\,s$^{-1}$, the wind takes $\sim$3000 years to reach a 10$\arcsec$ radius; we therefore assume implicitly that the dynamics of expansion has not changed significantly during the past 3000 years or so.

\begin{figure}
   \centering
   \includegraphics[width=0.9\textwidth]{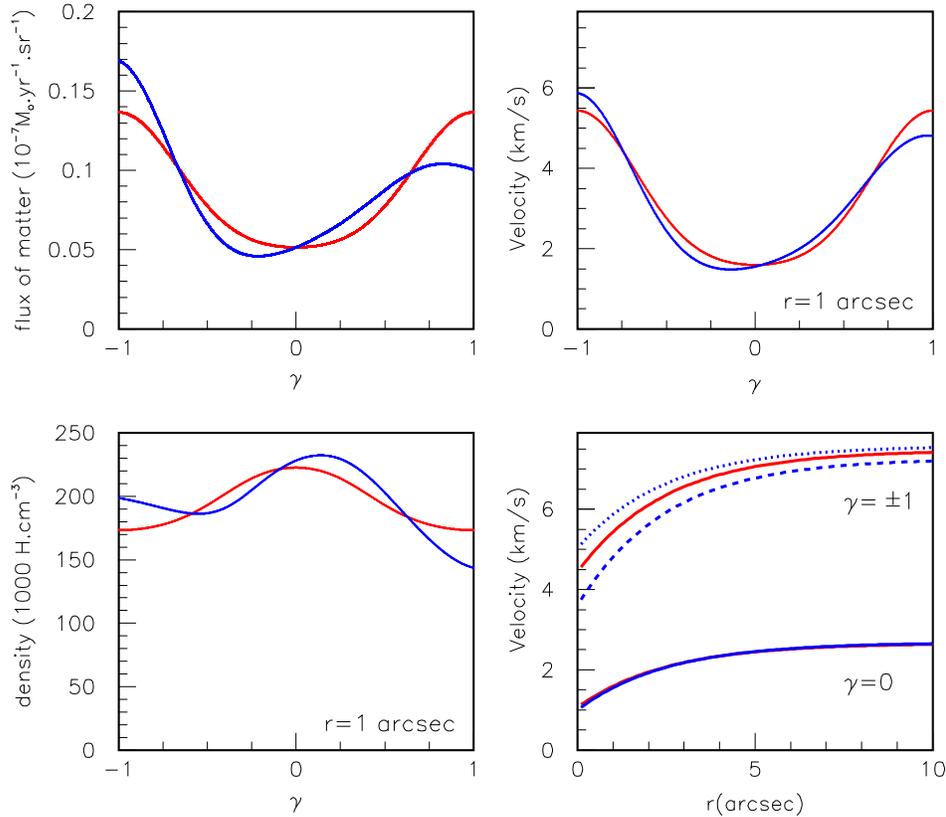}
   \caption{Dependence on the sine of the star latitude, $\gamma$, of the flux of matter (\textit{up-left}), of the wind velocity at $r=$\,1$\arcsec$ (\textit{up-right}) and of the gas density at $r=$\,1$\arcsec$ (\textit{down-left}). \textit{Down-right}: $r$-dependence of the equatorial (\textit{lower curves}) and polar (\textit{upper curves}) velocities. The best fit results of the standard model are shown in red and those of its modified asymmetric version in blue. The dashed curve is for $\gamma=$\,1 (\textit{north}) and the dotted curve for $\gamma=-$\,1 (\textit{south}).}
       \label{Fig2}
 \end{figure}

Figure~\ref{Fig2} displays the dependence over star latitude of the wind velocity, density and flux of matter as well as the $r$-dependence of the wind velocity.

%Fig 2
In comparison with Hoai et al.~(\cite{hoai}), two minor modifications have been introduced in the standard model, one related to the velocity gradients and the other to the radial distribution of the gas temperature. The aim is not to find the best possible analytic forms of $V(\gamma,r)$ and $T(r)$, but to have enough flexibility in the parameterization to understand how strongly the data constrain these quantities in different regions of $r$. 

\begin{figure}
   \centering
   \includegraphics[width=0.7\textwidth]{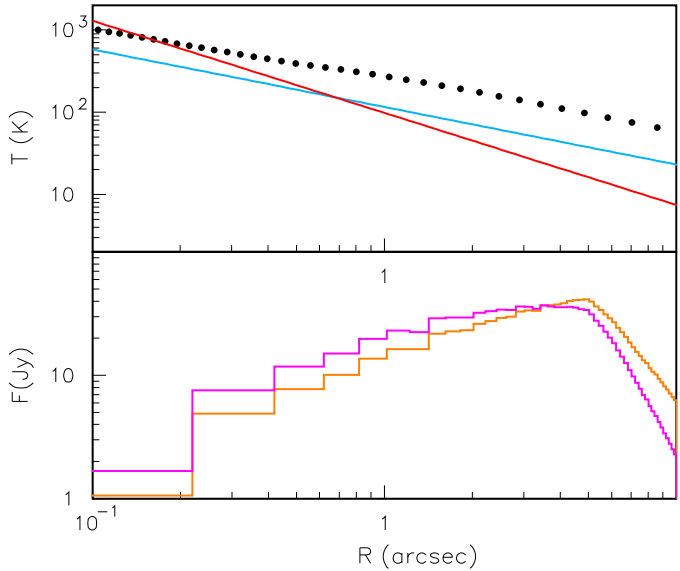}
   \caption{Dependence on $r$ of the gas temperature (\textit{upper panel}) and of the detected flux density obtained from the model (Jy per bin of 0.2 arcsec, \textit{lower panel}), in log-log scales. Temperatures are displayed for the present version of the standard model (\textit{red}), the version used in Hoai et al.~(\cite{hoai}, \textit{cyan}) and the radiative transfer calculations of Sch\"{o}ier and Olofsson~(\cite{scho}, \textit{black dotted line}). The flux densities are shown for CO(1-0) (\textit{orange}) and CO(2-1) (\textit{magenta}, divided by 10) separately. The restriction of the study to the 49 central cells implies an effective progressive truncation of the $r$ distribution from $\sim$5$\arcsec$ onward.}
   \label{Fig3}
 \end{figure}

For the velocity gradients, the use of a power law in Hoai et al.~(\cite{hoai}) has the disadvantage of suggesting that velocities keep increasing indefinitely, which is unphysical. Of course, such a suggestion is unfounded, the parameterization applying only to the region probed by the data, typically 0.2$\arcsec$ to 10$\arcsec$ (see lower panel of Figure~\ref{Fig3}). The form adopted here, an exponentially decreasing gradient, with velocities reaching a plateau, was preferred for this reason. It does not introduce additional parameters, simply replacing two power law indices by two exponential amplitudes, $\lambda_1$ and $\lambda_2$. Attempts to make the velocities reach a plateau at reasonable distances from the star, not exceeding 1$\arcsec$, have failed. As it is difficult to imagine a mechanism that would make it possible to maintain acceleration at large distances from the star, this result is probably the manner by which the model is able to best mimic a more complex reality. For example, simply artificially smearing the velocity distributions can remove the need for extended velocity gradients. Smearing in the model combines a thermal broadening of 0.02$\sqrt{T}$ km\,s$^{-1}$ and an \mbox{\textit{ad hoc}} instrumental smearing of 0.2 km\,s$^{-1}$; we see no justification to increase it significantly.

For the temperature, the approach in Hoai et al.~(\cite{hoai}) was to scale down, by a factor $\sim$2, the radial dependence proposed by Sch\"{o}ier and Olofsson~(\cite{scho}) from Monte Carlo simulations of spherical expanding shells, in order to match the luminosity and effective temperature of RS\,Cnc (Dumm and Schild~\cite{dumm}, Dyck et al.~\cite{dyck}, Perrin et al.~\cite{perr}). Here, by giving particular attention to the radial dependence of the ratio of the CO(2-1) and CO(1-0) emissions, we aim at a better understanding of the constraint imposed by the data on $T(r)$. In principle, both the velocity gradients and the radial dependence of the temperature control the $R_{yz}$ dependence of the ratio of CO(2-1) and CO(1-0) emissions. In practice, however, the velocity gradients are strongly constrained by the wind velocities, leaving the temperature as the main handle to best fit the $R_{yz}$ dependence of the emission ratio.

Several possible parameterizations of $T(r)$ have been tried and found adequate but none was able to improve the quality of the fit at $r=$\,0 (Figure~\ref{Fig1}). The $\sim$25$\%$ disagreement between model and observations in the centre of the spectral map, while probably significant, has therefore another cause than a wrong temperature, possibly turbulences and a lack of thermal equilibrium.

Figure~\ref{Fig3} displays the $r$-dependence of the gas temperature as obtained from the standard model best fit together with that used in Hoai et al.~(\cite{hoai}).  The standard model result uses as parameterization a power law $T(r)=T(1\arcsec)r^{-\alpha}$ with $T(1\arcsec)$ and $\alpha$ being two adjustable parameters.  Many different parameterizations have been explored with the aim of understanding what the data impose on $T(r)$. The result of this exploration can be summarized as follows. The temperature must reach a low value of $\sim$10$\pm$2 K at $r=$\,10$\arcsec$. The temperature gradient may then be small in absolute value at larger values of $r$, corresponding to a power law index of $-$0.7 or even less in absolute value. However, as soon as $r$ decreases, the temperature gradient must increase in absolute value and the $r$-dependence of the temperature must become accordingly steeper, with a power law index reaching values of the order of $-$1 at $r\sim$2$\arcsec$ to 5$\arcsec$ and of $-$1.1$\pm$0.1 at $r=$\,0.1$\arcsec$. The low value of the gas temperature at $r\sim$10$\arcsec$ is a cause of concern as the detection of HI at larger values of $r$ (Hoai et al.~\cite{hoai}) implies temperatures well in excess of 10 K. The radiative transfer calculation of Sch\"{o}ier \& Olofsson~(\cite{scho}) is expected to be more reliable at large values of $r$ than near the star, suggesting a power law radial dependence of the temperature with index $\sim-$\,0.7 beyond 10$\arcsec$. In the absence of reheating between 10$\arcsec$ and 30$\arcsec$, one would then expect the gas temperature at $r\sim$10$\arcsec$ to be above 30 K or so. Evidence against higher temperatures is given in Figure~\ref{Fig1} (right panel) where the $r$-dependence used in Hoai et al.~(\cite{hoai}) is seen to imply too high a \mbox{$\langle$\textit{CO(2-1)}$\rangle$\textit{/}$\langle$\textit{CO(1-0)}$\rangle$} ratio. We note that large values of $R_{yz}$ receive contributions from large values of $r$ exclusively while small values of $R_{yz}$ receive contributions from the whole spectrum of the $r$ distribution. As the dependence on temperature of the ratio of the populations of the $J=$\,2 and $J=$\,1 states is maximal around 10 K, it is not surprising that the data constrain $T(r)$ in the large $r$ region, where the temperature takes values in such a range. At small values of $r$, anyhow, the temperature is large enough for both rotational levels to be maximally populated. This implies that even at small $R_{yz}$ values the \mbox{$\langle$\textit{CO(2-1)}$\rangle$\textit{/}$\langle$\textit{CO(1-0)}$\rangle$} ratio is mostly governed by the temperatures reached at large $r$ values.

%Fig 3
The evidence in favour of low temperatures in the $r\sim$10$\arcsec$ region and below raises the question of the dependence of the temperature on star latitude. This was studied by using in the model forms of $T(r)$ allowing for it to depend on the star latitude. Various forms have been tried, all giving very similar fits to the 49 central velocity spectra. For example, a power law with index $\alpha_0$ up to $r=$\,1$\arcsec$ and with index $\alpha_1(1+\alpha_2|\gamma|)$ for $r>$1$\arcsec$ reduces the value of $\chi^2$ by 3$\%$ and gives temperature values of 1064 K for $r=$\,0.1$\arcsec$, 97 K for $r=$\,1$\arcsec$, 9.7 K for $r=$\,10$\arcsec$ at the equator and 6.3 K for $r=$\,10$\arcsec$ at the poles. In all cases, at $r=$\,10$\arcsec$, the temperature reaches $\sim$10 K at the equator and $\sim$6 K at the poles. It is directly constrained by the observed map of the \mbox{\textit{CO(2-1)/CO(1-0)}} ratio at large $R_{yz}$ values. While the consistently lower value of the polar temperature suggests that the associated more rapid expansion plays a cooling role, the overall temperature remains low at all latitudes and does not remove the concern expressed earlier in relation with the observation of HI emission further out.

%Fig 4
 \begin{figure}
   \centering
   \includegraphics[width=0.9\textwidth]{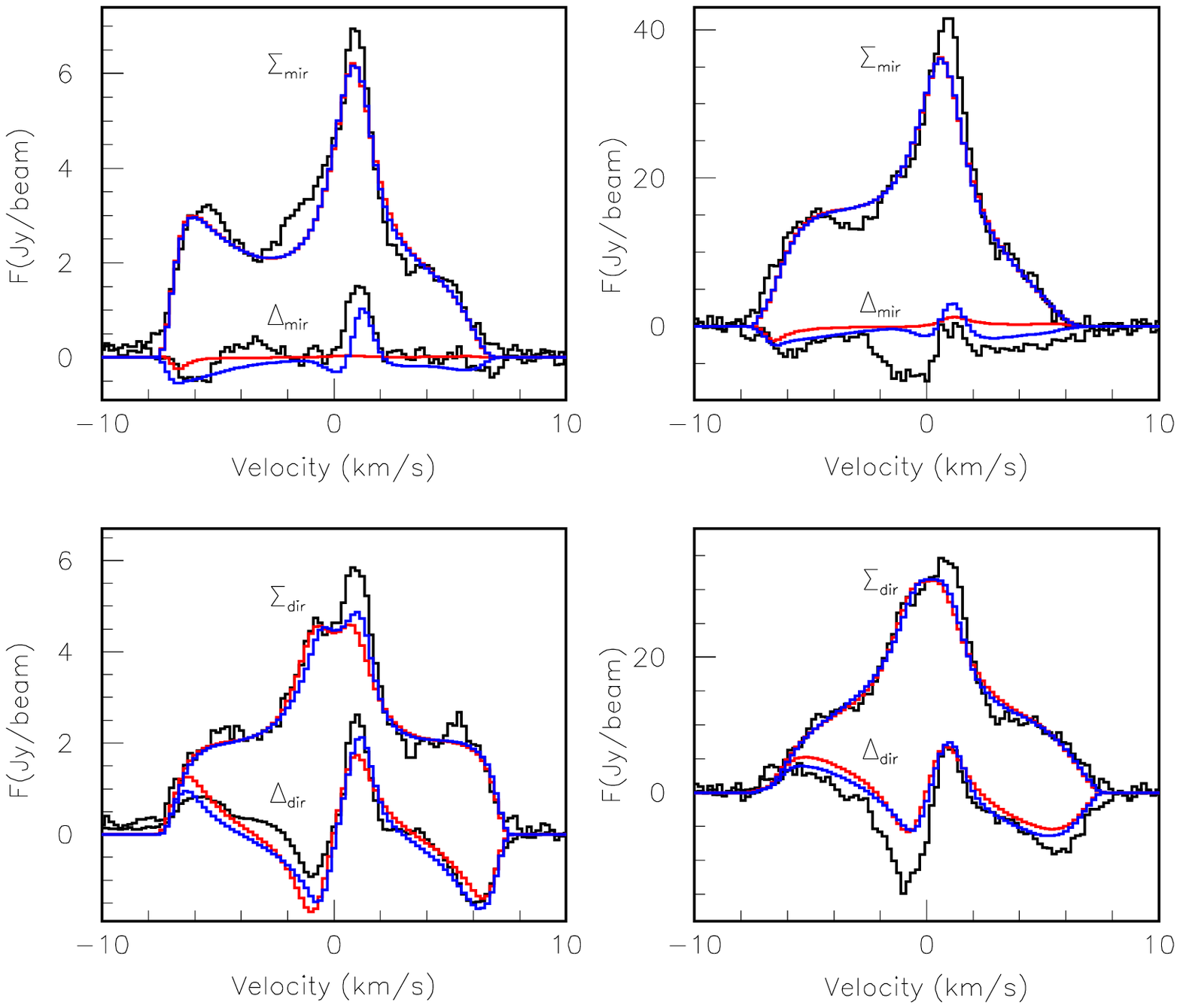}
   \caption{Velocity distributions of $\Sigma_{dir}$, $\Delta_{dir}$, $\Sigma_{mir}$, $\Delta_{mir}$, evaluated over the 24 pairs of diametrically opposite spectra of the CO(1-0) (\textit{left}) and CO(2-1) (\textit{right}) spectral maps. The upper panels are for mirror quantities and the lower panels for direct quantities. In each case, the data are shown in black, the best fit results of the standard model in red and of its modified asymmetric version in blue. The reference velocity is 7.25 km\,s$^{-1}$, the CO(1-0) data of Libert et al.~(\cite{libe}) and Hoai et al.~(\cite{hoai}) having been corrected by +0.5 km\,s$^{-1}$.}
   \label{Fig4}
 \end{figure}
 \section{Deviation from central symmetry in CO(1-0) and CO(2-1) emission}

Having now adjusted the parameters of the standard model to best fit the set of reprocessed data, we are in a position to study to which extent they obey central symmetry. In the standard model, the only source of central asymmetry is the presence of absorption, which is found to have a nearly negligible effect. The velocity distributions of $\Sigma_{dir}$, $\Delta_{dir}$, $\Sigma_{mir}$ and $\Delta_{mir}$, evaluated over the northern half (24 spectra) of the CO(1-0) and CO(2-1) data are displayed separately in Figure~\ref{Fig4}. Significant asymmetries are evidenced with $2\Delta_{mir}/\Sigma_{mir}$ reaching up to $\sim$30$\%$ in some cases but being usually well below a few percent. They display important regularities that can be schematically summarized as a dominance of two features: a narrow low velocity spike between $\sim$0 and $\sim$2 km\,s$^{-1}$ in the northern part of the sky (meaning positive values of $\Delta_{mir}$), particularly strong in the CO(1-0) data; and a broad velocity span in the southern part (meaning negative values of $\Delta_{mir}$), particularly strong in the CO(2-1) data. As expected, the asymmetries calculated in the standard model (resulting from absorption exclusively) are seen to be negligible. We also note that the asymmetries arising from the misalignment of the image due to the proper motion of the star between year 2000 and the times when observations were made amount to typically a tenth of the asymmetries observed in the reprocessed data. This artefact is absent from the reprocessed data.

% Table 1
\begin{table}
\caption{Best fit model parameters with and without explicit asymmetries. Values in parentheses display the last digits corresponding to an increase of $\chi^2$ of 1$\%$ when the parameter is varied with the others kept constant. The distance to the star is taken to be 143 pc (van Leeuven~\cite{vanl}).}             
\label{Table1}      
\centering        
\renewcommand{\arraystretch}{1.5}
\resizebox{12cm}{!} {
\begin{tabular}{|c|c|c|}
\hline 
Parameter & \makecell{Best fit value symmetric \\ (standard model)} & \makecell{Best fit value asymmetric \\ (modified standard model)} \\
%          & (standard model) & (modified standard model)\\
\hline
$AI (^\circ)$ & 52 (2)& 53 (2)\\
\hline
$PA (^\circ)$ & 9 (6) & 8 (5)\\
\hline
$\sigma$ & 0.33 (2) & 0.33 (2)\\
\hline
$V_1$ (km\,s$^{-1}$) & 4.91 (15) & 4.83 (11)\\
\hline
$V_2$ (km\,s$^{-1}$) & 2.56 (9) & 2.58 (9)\\
\hline
$\lambda_1$ & 0.36 (11) & 0.33 (10)\\
\hline
$\lambda_2$ & 0.62 (7) & 0.66 (7)\\
\hline
$\dot{M}_1$ (10$^{-7}$ M$_\odot$/yr) & 1.09 (13) & 1.07 (11)\\
\hline
$\dot{M}_2$ (10$^{-7}$ M$_\odot$/yr) & 0.62 (5) & 0.64 (5)\\
\hline
$T$ (1$\arcsec$) (K) & 98 (6) & 99 (6)\\
\hline
$\alpha$ & 1.12 (5) & 1.11 (5)\\
\hline
$\varepsilon_{\dot{M}_1}$ & 0 & $-$0.96 (15)\\
\hline
$\varepsilon_{\dot{M}_2}$ & 0 & 0.95 (16)\\
\hline
$\varepsilon_{V_1}$ & 0 & $-$0.40 (3)\\
\hline
$\varepsilon_{V_2}$ & 0 & 0.70 (6)\\
\hline
$\chi^2/dof$ & 1.240 & 1.137\\
\hline
\end{tabular} }
\end{table}
In order to improve the agreement between observations and best fit results of the model, the latter must be modified to make room for the observed central asymmetries. How to do so is somewhat arbitrary. A possibility that has been explored is to assume the presence of a cold cloud in the northern part of the sky that mimics the northern excess observed in the CO(1-0) data. Here we prefer to retain a more phenomenological approach, which does not presume of the physics nature of the effect. As we need to have a northern excess confined to a low velocity spike and a southern excess covering a broad velocity range, we simply replace the parameters $\dot{M}_1$, $\dot{M}_2$, $V_1$ and $V_2$, by  $(1+\varepsilon_{\dot{M}_1}\gamma)\dot{M}_1$, $(1+\varepsilon_{\dot{M}_2}\gamma)\dot{M}_2$, $(1+\varepsilon_{V_1}\gamma)V_1$ and $(1+\varepsilon_{V_2}\gamma)V_2$. The result is a significant decrease of the value of $\chi^2$ with important asymmetries introduced in the $\gamma$ dependences of the parameters, as displayed in Figure~\ref{Fig2}. The values taken by the former parameters do not change much (Table \ref{Table1}) but the values obtained for the $\varepsilon$ asymmetry parameters are $\varepsilon_{\dot{M}_1}=-$\,0.96, $\varepsilon_{\dot{M}_2}=$\,0.95, $\varepsilon_{V_1}=-$\,0.40 and $\varepsilon_{V_2}=$\,0.70. The $\chi^2$ minimization uses uncertainties combining a 9$\%$ [8$\%$] error with a 14 mJy [116 mJy] noise for CO(1-0) [CO(2-1)] respectively, adjusted to have for each set of observations a value close to the number of degrees of freedom.

\begin{figure}
   \centering
   \includegraphics[width=.95\textwidth]{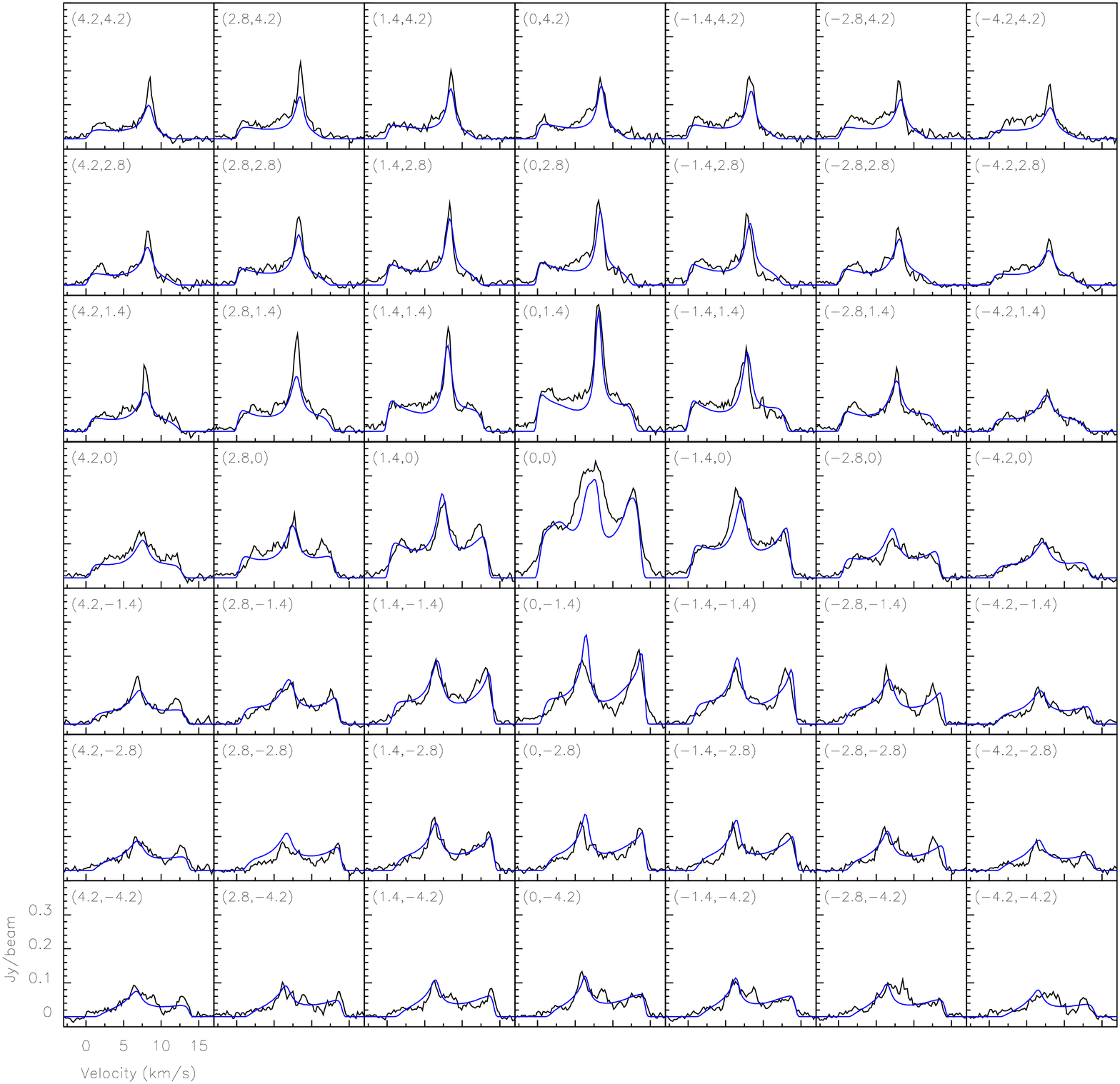}
   \caption{Spectral map centred on the star of the CO(1-0) observations (\textit{black}) and best fit results of the modified asymmetric version of the standard model (\textit{blue}). Steps in right ascension and declination are 1.4$\arcsec$. The synthesized circular beams are Gaussian with a full width at half maximum of 1.2$\arcsec$. The coordinates of the beam centre are indicated in each cell.}
   \label{Fig5}
 \end{figure}

\begin{figure}
   \centering
   \includegraphics[width=.95\textwidth]{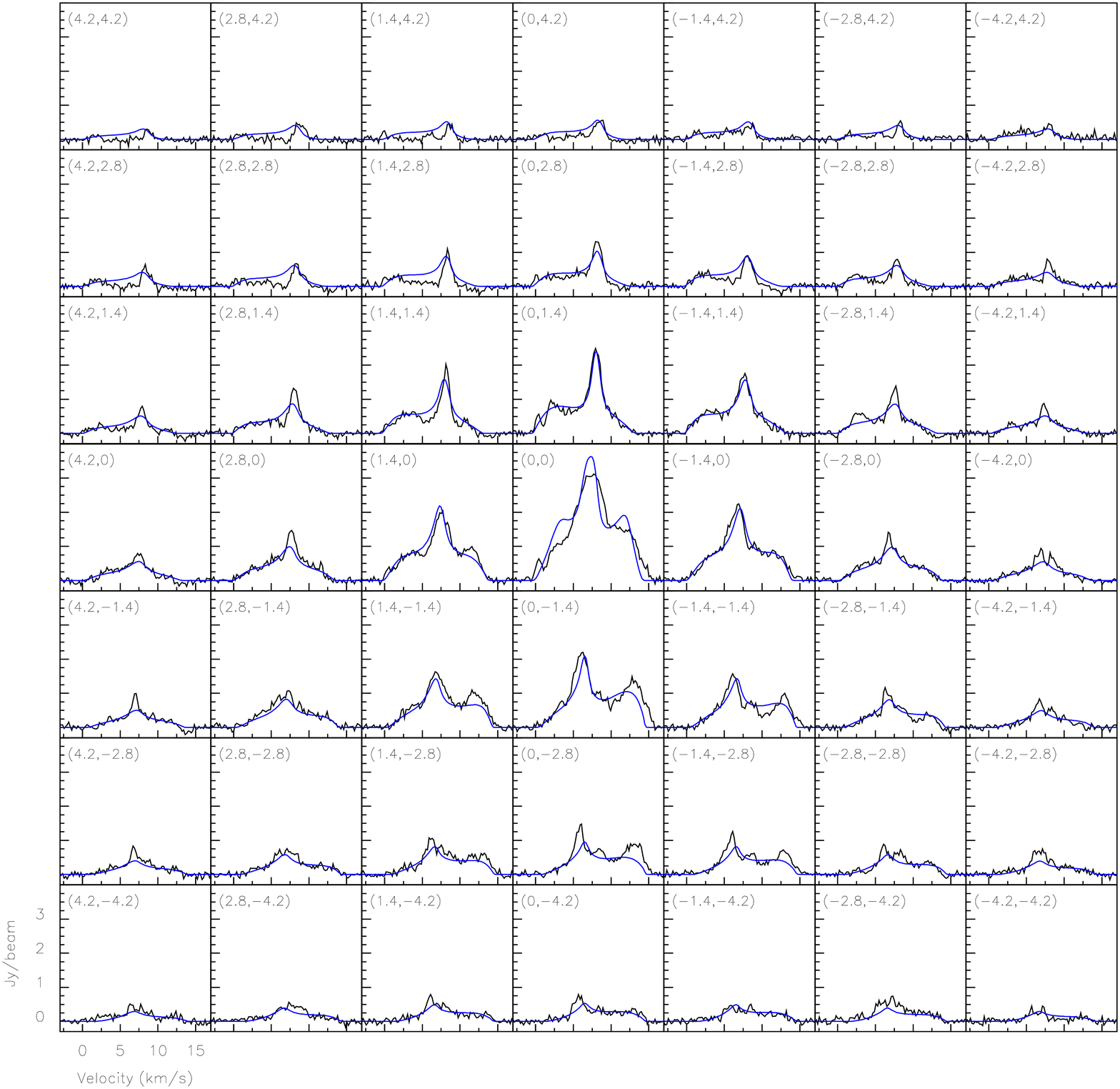}
   \caption{Spectral map centred on the star of the CO(2-1) observations (\textit{black}) and best fit results of the modified asymmetric version of the standard model (\textit{blue}). Steps in right ascension and declination are 1.4$\arcsec$. The synthesized circular beams are Gaussian with a full width at half maximum of 1.2$\arcsec$. The coordinates of the beam centre are indicated in each cell.}
   \label{Fig6}
 \end{figure}

The asymmetries obtained from the modified model are illustrated in Figure~\ref{Fig4}. While qualitatively in good agreement with observations, they fail to reproduce them quantitatively. However, lacking a clear physical interpretation of the observed asymmetries, we did not attempt fine tuning the model in order to improve the agreement with observations. Model and observations are compared in Figures~\ref{Fig5} and~\ref{Fig6}. We have checked that the new asymmetric model, using the values of the parameters optimized on the 49 central spectra, gives also good fits when extending the spectral maps to the 9$\times$9=81 central spectra.

%Fig 5
%

%Fig 6
\section{Conclusion} 

The detailed study of CO(1-0) and CO(2-1) emission from the circumstellar envelope of the AGB star RS\,Cnc has revealed departures from central symmetry that turned out to be efficient tools for the exploration of some of its properties. In a first phase, they provided evidence for offsets in velocity, declination and right ascension that were inherent to the procedure of data reduction and were causing small biases in the data. These have been removed by reprocessing the data with the spectral map centred on the star position at the time when the observations were made and the velocity spectra centred on the star Doppler velocity. 

The reprocessed data allow for a detailed comparison of CO(1-0) and CO(2-1) emissions that reveals the need for the gas to reach lower temperatures than expected over the radial range probed by CO emission. The observation of HI emission further out seems therefore to require some reheating. Evidence for lower gas temperatures near the poles than at the equator has been presented, probably associated with the associated more rapid expansion.

It is also noteworthy that the density profile is almost spherically symmetric (Fig~\ref{Fig2}, and Fig~\ref{Fig6} in Hoai et al.~\cite{hoai}). In RS Cnc, the asymmetry seems mainly a kinematical effect. Possibly, RS Cnc is in an early stage of the development of the axi-symmetry that leads to a bipolar outflow. The flux of matter occurs preferentially in the polar directions, which on larger scales than probed in CO is confirmed by the elongation of the HI central source reported by Hoai et al.~(\cite{hoai}). The results of our modelling do not favour current models invoking magnetic fields (Matt et al.~\cite{matt}) or stellar rotation (Dorfi \& H\"ofner~\cite{dorf}) for inducing an axi-symmetrical distribution of the matter in the circumstellar environment. This may plead in favour of the binary hypothesis but presently we have no proof of it.

The study of the deviation from central symmetry of the wind morphology and kinematics has revealed the presence of significant central asymmetries that can be schematically summarized as a northern excess confined to a narrow spike of low velocity, mostly in CO(1-0) data, and a southern excess covering a broad velocity range, mostly in CO(2-1) data. A simple parameterization of the asymmetry has been suggested, providing a good qualitative description of the main features but failing to give a precise quantitative account. Lacking a clear physical interpretation of the observed asymmetries, we did not attempt refining the model. The observation of significant departures from central symmetry is in itself an important result in the context of the transition from the AGB to post-AGB and Planetary Nebulae phases that are known to display growing asymmetries, the origin of which is not clearly understood. From the present work, it should be traced to the inner part of the envelope where observations with higher spatial resolution are still needed.

A possible source of central asymmetry is the suggested existence of a companion source located $\sim$0.98$\arcsec$ West and  $\sim$0.63$\arcsec$ North of the main source and accreting gas from its wind (Hoai et al.~\cite{hoai}). At the present stage such an interpretation is only tentative, it is not even possible to discriminate between a compact source and a clump of gas. While the main source is seen on both line and continuum maps, the companion source is not seen in the continuum but only on the line. However, the large distance of the companion from the main star, $\sim$170 a.u. (implying an orbital period of some 2200 years), makes it unlikely that it could play an important role in the shaping of the bipolar outflow. When binaries are invoked as triggers of the circumstellar envelope bipolar morphology, the companions are much closer with orbital periods of the order of a year (Jorissen et al.~\cite{jori}; van Winckel et al.~\cite{vanw}). In the case of wide binaries, the companion plays a lesser role but is still likely to accrete gas and induce a weak asymmetry or produce structures such as spirals or detached shells (Maercker et al.~\cite{maer}; Perets \& Kenyon~\cite{pere}). Of course, one cannot exclude the presence of an unobserved close companion playing an important role in shaping the circumstellar envelope of RS\,Cnc, but this is of no relevance to the present discussion. Moreover, the interpretation of the observed morphology as due to a companion star is by no means certain as clumps and knotty jets have been observed in late AGB and post-AGB phases (for recent references, see Balick et al.~\cite{bali}).

\acknowledgements{
This work was done in the wake of an earlier study, the results of which have been published (Hoai et al.~\cite{hoai}). We thank all those who contributed to it initially, in the phases of design, observations and data reduction. We are grateful to Dr Lynn Matthews for her constant interest and helpful comments. Financial support from the World Laboratory, from the French CNRS in the form of the Vietnam/IN2P3 LIA and the ASA projects, from FVPPL, from PCMI, from the Institute for Nuclear Science and Technology, from the Vietnam National Foundation for Science and Technology Development (NAFOSTED) under grant number 103.08-2012.34 and from the Rencontres du Vietnam is gratefully acknowledged. One of us (DTH) acknowledges support from the French Embassy in Ha Noi. This research has made use of the SIMBAD and ADS databases.}

%(Clemens~\cite{clem85})  

\label{lastpage}

\end{document}